\documentclass[pre,showpacs,nofootinbib]{revtex4-1}
\usepackage{amssymb,amsmath}
\usepackage{natbib}
\usepackage{graphicx}
\usepackage{color}
\usepackage{bm}

	 \renewcommand{\vec}[1]{\bm{#1}}

\begin{document}

\title{Equilibration of Quasi-Integrable Systems}

\author{Tomer Goldfriend}
\email{tomergf@gmail.com}
\affiliation{Laboratoire  de  Physique  Statistique, D\'epartement de physique de l’ENS, \'Ecole  Normale  Sup\'erieure,
PSL  Research  University;  Universit\'e  Paris  Diderot, Sorbonne  Paris-Cit\'e;  Sorbonne
Universit\'es,  UPMC  Univ.  Paris  06,  CNRS;  24  rue  Lhomond,  75005  Paris,  France}

\author{Jorge Kurchan} 
\affiliation{Laboratoire  de  Physique  Statistique, D\'epartement de physique de l’ENS, \'Ecole  Normale  Sup\'erieure,
PSL  Research  University;  Universit\'e  Paris  Diderot, Sorbonne  Paris-Cit\'e;  Sorbonne
Universit\'es,  UPMC  Univ.  Paris  06,  CNRS;  24  rue  Lhomond,  75005  Paris,  France}

\date{\today}

\begin{abstract}
We study the slow relaxation of isolated quasi-integrable systems, focusing on the classical problem of Fermi-Pasta-Ulam-Tsingou (FPU) chain. It is well-known that the initial energy sharing between different linear-modes can be inferred by the integrable Toda chain. Using numerical simulations, we show explicitly how the relaxation of the FPU chain toward equilibration is determined by a slow drift within the space of Toda's integrals of motion. We analyze the whole spectrum of Toda-modes and show how they dictate, via a Generalized Gibbs Ensemble (GGE), the quasi-static states along the FPU evolution. This picture is  employed to devise a fast numerical integration, which can be generalized to other quasi-integrable models. In addition, the GGE description leads to a fluctuation theorem, describing the large deviations as the system flows in the entropy landscape.
\end{abstract}

\maketitle

\section{Introduction}
\label{sec:intro}

The equilibration of an isolated macroscopic system has been studied from the beginning of Thermodynamics, e.g., the discussions between Boltzmann and Loschmidt~\cite{LoschmidtEcho_Schol}, up to recent experimental studies on isolated quantum systems~\cite{Langen_etal2016}. Integrable systems do not relax to the usual Gibbs measure, as they posses many symmetries, restricting the dynamics to high-dimensional manifold. While integrable models are rare, many physical systems, such as planetary systems~\cite{Morbidelli} or small vibrations within an elastic plate~\cite{During_etal2006}, can be considered as quasi-integrable---  described by an integrable Hamiltonian which is weakly perturbed. In low-dimensional systems, the Kolmogorov-Arnold-Moser (KAM) theorem guarantees that some of the invariant tori are stable against integrability-breaking interactions, or, according to Nekhoroshev theorem, can change exponentially in time. For macroscopic systems, these effects are generically vanishingly small with the perturbation, and thus irrelevant.

When an isolated quasi-integrable system is initially excited to an atypical state, it is expected to relax slowly toward equilibration. This was the subject  of one of the first numerical simulations in Theoretical Physics--- the pioneering study by Fermi, Pasta, Ulam, and Tsingou (FPU)--- on which we focus in the current paper. (For the significant contribution of Mary Tsingou see Ref.~\cite{Dauxois2008}.) Typically, quasi-integrable systems are chaotic, having a positive Lyapunov exponent, a quantity which measures the exponential separation between initially nearby trajectories. However, in such systems the corresponding Lyapunov time is usually much shorter than the characteristic time of diffusion within the space of quasi-conserved quantities. For example, a FPU chain of size $N=1024$ and energy density $\epsilon=E/N=10^{-4}$ has a Lyapunov time of $10^6$, much shorter than the relaxation time of $10^{10}$~\cite{Benettin_etal2013,Benettin_etal2018}. Another example is the Solar System, there, the Lyapunov time is 5-10 Myrs, whereas the stability time is over  5 Gyrs~\cite{Laskar2008}. In fact, we may refer to this generic feature of time-scale separation as a definition for `quasi' - integrability.  

The characteristics of quasi-integrable systems can be captured by considering a simple model of an integrable system perturbed by weak stochastic noise, as studied in Ref.~\cite{Lam&Kurchan2014}. There, it was shown that weak noise induces chaotic motion confined primarily within the invariant tori, thus allowing the system to diffuse quasi-statically `from torus to torus'. Furthermore, there is no KAM regime for any amplitude of the stochastic perturbation. Here, we use this picture as a motivation for understanding the FPU dynamics. We do not, however, prove that a small integrability-breaking perturbation is equivalent to a stochastic one, although discussion on the resemblance of the two situations is given at the end of the current paper. 

Since the original work by Fermi, Pasta, Ulam and Tsingou, an extensive study has been dedicated to explore several physical processes in this  nonlinear chain; see Refs.~\cite{Berman&Izrailev2005,Gallavotti,Dauxois&Ruffo2008} for comprehensive reviews. Previous works established the intimate relation between the integrable Toda chain and the quasi-integrable FPU one. This was done along the two aspects: ({\it i}) The FPU dynamics involves solitons--- long-lived localized excitations. It was initially found by using the connection between the FPU model, limited to the hydrodynamic limit, and the integrable KdV equation~\cite{Zabusky&Kruskal1965}. Later, a relation to the Toda chain was shown as well~\cite{Toda}. ({\it ii})  At short timescales the state of the FPU chain is completely determined by {\it fully-explored} Toda tori~\cite{Ferguson_etal1982,Benettin_etal2009,Benettin_etal2013}; see also Fig.~\ref{fig:BenettinLike}(a) below. 

In the current paper we address the role of quasi-integrability in the slow  dynamics of the FPU chain toward equipartition. The expected adiabatic change of the Toda conserved quantities under the FPU dynamics was pointed out in Ref.~\cite{Ferguson_etal1982}, and recent works studied the slow evolution of specific Toda modes and their time correlations~\cite{Ponno_etal2011,Benettin_etal2013}. In the case of other quasi-integrable models, a hydrodynamic description of weakly conserved quantities was used to address, e.g., energy cascade in weakly coupled wave modes (weak turbulence)~\cite{Zakharov&Filonenko1967,During_etal2006}, or slow relaxation in a gas with long-range interactions~\cite{Chavanis2007}. Here, we explore in detail how the full spectrum of Toda conserved quantities dictate the evolution of the FPU chain. The analysis of both models involves intractable computations, and we resort to numerical simulations. In Sec.~\ref{sec:Benettin}, after setting the main equations, we start with presenting a clear evidence of the Toda quasi-static states along the FPU slow dynamics. In Sec.~\ref{sec:TodaJ} we define the Toda conserved quantities which are used along the paper, and introduce the notion of Generalized Gibbs Ensemble (GGE)--- a statistical measure of a system with several conserved quantities. This description suggests a fast numerical integration and leads to a fluctuation theorem, as discussed in Secs.~\ref{sec:numerical} and~\ref{sec:GXFT} respectively. The analysis involves several technical issues, whose details are given in the Appendix. A relation to other quasi-integrable systems, such as the Solar System dynamics, or integrable models perturbed by random noise are discussed in the Sec.~\ref{sec:Con}, where we conclude our results.

\section{quasi-static states in the FPU experiment}
\label{sec:Benettin}

\subsection{The FPU Hamiltonian and its close integrable systems}

We study a one dimensional chain of $N+1$ beads with a unit mass set to 1, which follows the FPU Hamiltonian
\begin{equation}
\mathcal{H}_{\rm FPU}=\frac{1}{2}\sum^N_{n=1} p_n^2 + \sum^N_{n=0} V_{\rm FPU} (q_{n+1}-q_{n}),\qquad V_{\rm FPU}(r)=\frac{1}{2}r^2+\alpha\frac{1}{3}r^3+\beta\frac{1}{4}r^4,
\label{eq:H_FPU}
\end{equation}
where the case of fixed-ends is considered, $q_0=q_{N+1}=0$. The energy density of the system is designated by $\epsilon=E/N$, where $E$ is the total energy of the chain. The behavior of the system depends on $N$, $\epsilon$, and the parameter $\beta$; the $\alpha$ parameter, unless is zero, can be rescaled by the energy density~\cite{Benettin_etal2013} and is set to $\alpha=1$ hereafter.

We focus on the case of small perturbations, i.e., $\epsilon \ll 1$. In this limit the FPU potential is a slight correction to the linear chain and the Toda chain: 

The {\it linear potential} 
\begin{equation}
V_{\rm L}(r)=\frac{1}{2}r^2     
\end{equation}
satisfies $|V_{\rm L}-V_{\rm FPU}|\sim \alpha r^3\sim \alpha \sqrt{\epsilon}$. 
It corresponds to the normal (Fourier) modes
\begin{equation}
\begin{pmatrix}
Q_k\\P_k	
\end{pmatrix}
=\sqrt{\frac{2}{N+1}}\sum_{n=1}^N \begin{pmatrix}
q_n\\p_n	
\end{pmatrix}
\sin\left(\frac{\pi k n}{N+1}\right),
\end{equation}
and an energy spectrum
\begin{equation}
E_k=\frac{1}{2}(P_k^2+\omega_k^2 Q_k^2), \qquad \omega_k=2\sin\left(\frac{\pi k}{2(N+1)}\right).
\end{equation} 
The normal modes can be rotated while keeping the energy profile constant
\begin{equation}
\begin{pmatrix}
	\omega_k Q_k \\
	P_k
\end{pmatrix}\rightarrow
\begin{pmatrix}
	\cos\phi_k & -\sin\phi_k \\
	\sin\phi_k & \cos\phi_k
\end{pmatrix}
\begin{pmatrix}
	\omega_k Q_k \\
	P_k
\end{pmatrix}.
\end{equation}
The pairs $(E_k,\phi_k)$ constitute a geometry of invariant tori for the linear chain. We shall call  {\it normal modes ensemble}: a set of configurations  in which a fraction of the Fourier modes are excited with randomly chosen angles $\phi_k$,
\begin{equation}
\left\{ \begin{array}{l l}
E_k=N\epsilon/K, &\quad 1\leq k\leq K\\
E_k=0,	&\quad K< k\leq N
\end{array} \right.,
\qquad0\leq \phi_k\leq 2\pi.
\label{eq:NME}
\end{equation}

{\it The Toda potential} reads:
\begin{equation}
V_{\rm Toda}(r)= V_0 \left(e^{A r}-1-2A r\right),
\label{eq:V_Toda}
\end{equation}
Taylor expansion, together with choosing the parameters $V_0=(2\alpha)^{-2}$ and $A=2\alpha$, implies $|V_{\rm Toda}-V_{\rm FPU}|\sim \beta r^4\sim \epsilon$. Thus, the Toda model is a better approximation to the FPU chain.  

\subsection{The FPU numerical experiment}

We preform the following numerical experiment: The system is initialized with a normal modes ensemble, by fixing $K=N/10$ and taking uniformly distributed $\phi_k$ according to Eq.~\eqref{eq:NME}. This corresponds to  excitations of a small fraction of the longest wave-lengths and allows us to study the statistical properties of the system~\cite{Carati_etal2007}. Then, we integrate the system with the FPU Hamiltonian, Eq.~\eqref{eq:H_FPU}. The nonlinear terms in the potential give rise to an energy sharing between different normal modes. Thus, one would expect that at long times the system reaches an equilibrium, which, given the weak coupling between Fourier modes, amounts to an approximate equipartition 
$$\lim_{t\to\infty}\langle E_k(t)\rangle \sim \epsilon,$$
Here, the average is taken over the initial conditions (i.e. the normal mode ensemble above)~\footnote{We note that, unless otherwise stated, we do not employ an additional time-average $1/t\int_0^t E_k(t')dt'$.}. 

We aim to explore the properties of the system in the {\it thermodynamic and hydrodynamic limits}. Accordingly, throughout the paper we focus on the specific case  $N=511$, $\epsilon=10^{-3}$, $\alpha=1$, $\beta=2$, and the initial ensemble is defined with $K=N/10$. Details of the numerical integration are given in the Appendix.

\subsection{quasi-static states}
\begin{figure}
\centerline{\resizebox{0.45\textwidth}{!}{\includegraphics{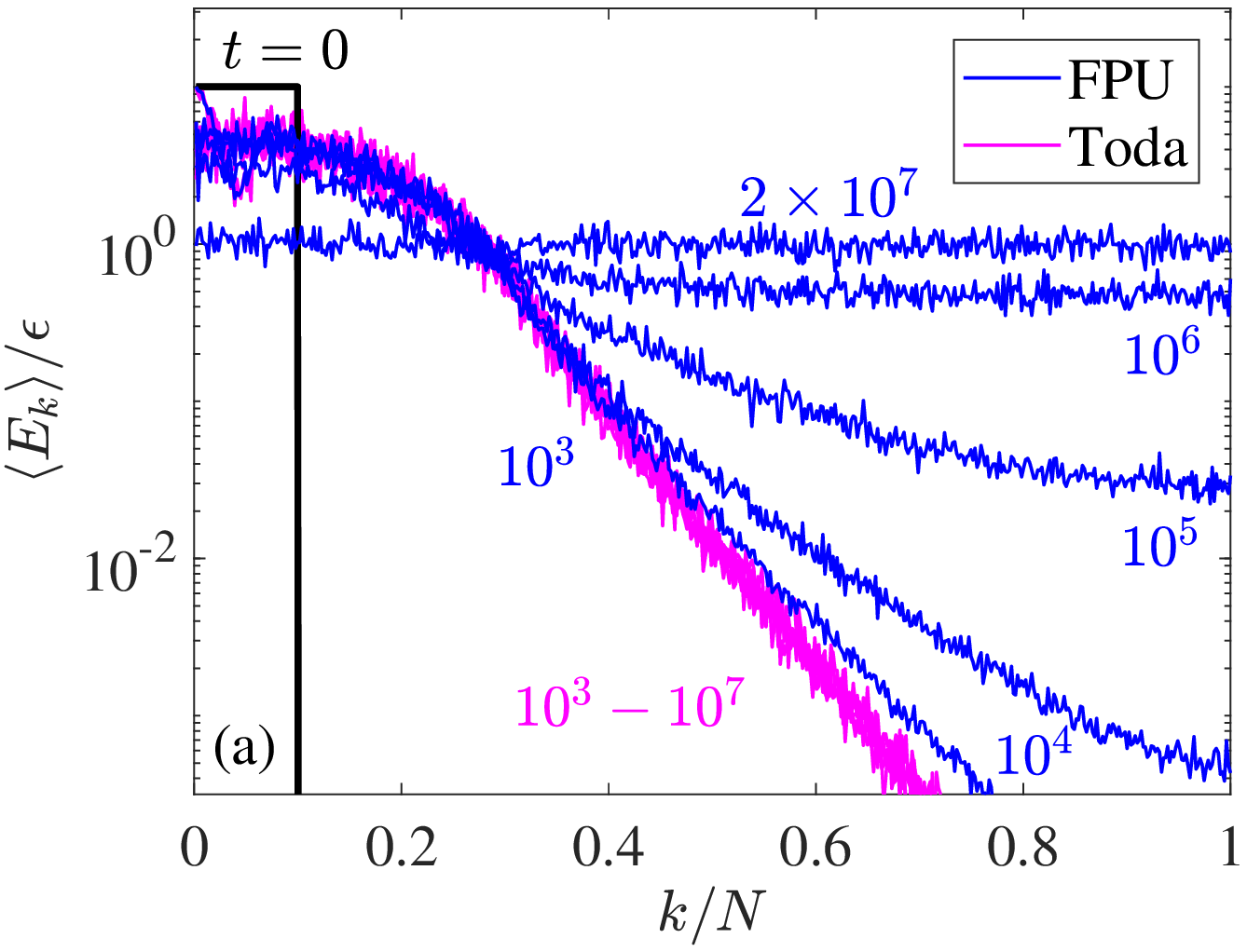}}
\hspace{1cm}
\resizebox{0.45\textwidth}{!}{\includegraphics{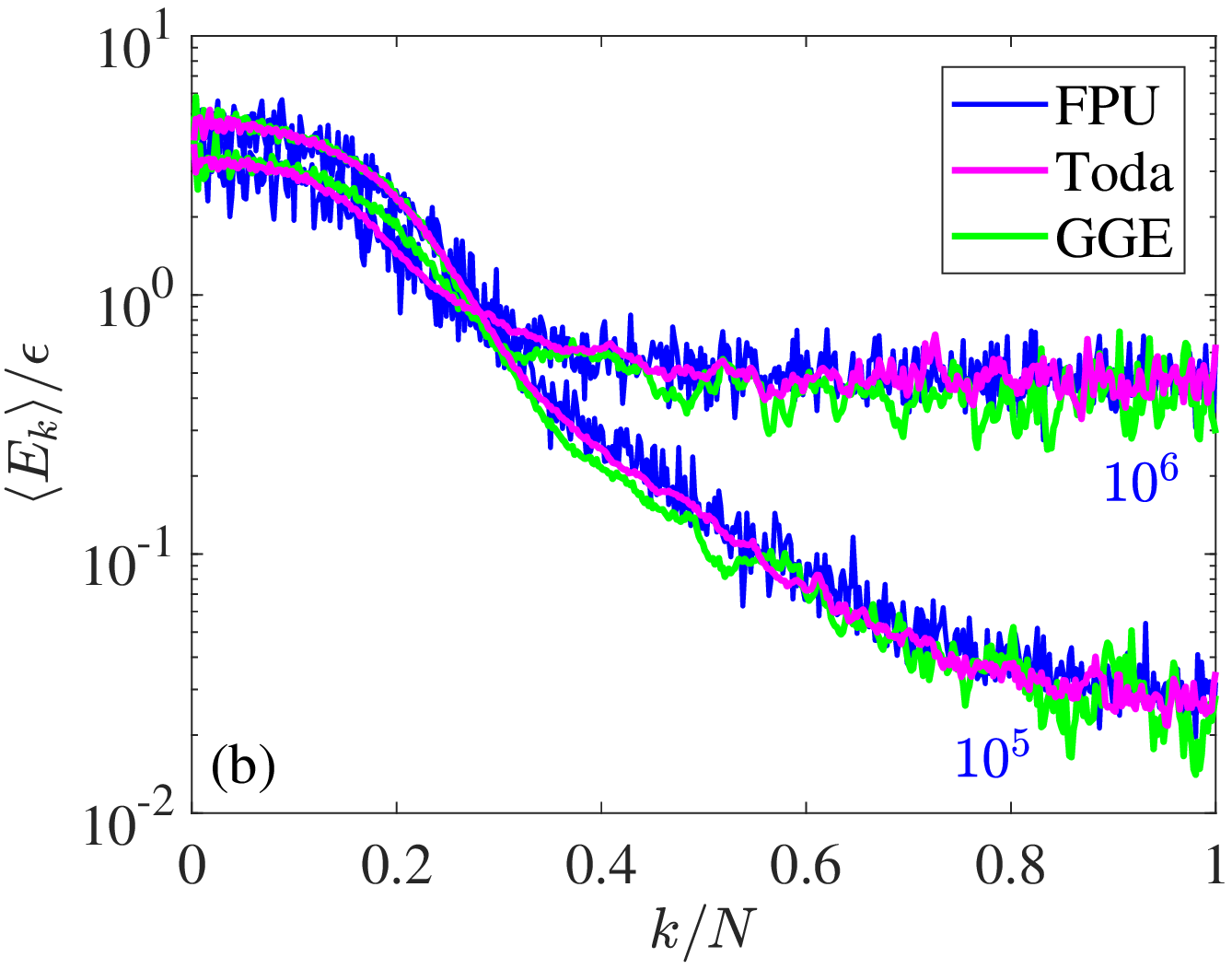}}}
\caption[]{The energy of the normal modes at different times for the FPU (blue) and the Toda (magenta) dynamics. The parameters of the system are specified in the text. (a) In both cases the integration starts with an initial ensemble at $t=0$ satisfying a step function profile (black line). The blue curves show the equilibration of the FPU chain, whereas the Toda dynamics reaches a terminal profile. The average is over 102 for the former, and over 31 out of these for the latter. (b) Here the blue curves are the ones of (a). The magenta curves correspond to initial conditions given at times $t=10^5$ and $t=10^6$ by the blue curves. The Toda profiles are averaged over time $\Delta t=10^6$ and 10 initial conditions. The green curves are obtained by a microcanonical GGE ensemble; see Sec.~\ref{ssec:GGE}.}
\label{fig:BenettinLike}
\end{figure}

In Figure \ref{fig:BenettinLike}(a) we show the slow relaxation of the system towards equipartition by plotting the profiles of $\langle E_k \rangle$ at different times (solid blue lines). Following the lines of Ref.~\cite{Benettin_etal2013} we also integrate the initial ensemble with the integrable Toda potential, Eq.~\eqref{eq:V_Toda}. Here, the system reaches a terminal, non-ergodized state. This corresponds to  the long-lived initial profile that was observed in the original work by Fermi, Pasta, Ulam, and Tsingou. 

Following the rationale mentioned in the Introduction, one would expect that each blue curve in Fig.~\ref{fig:BenettinLike}(a) stands for some Toda tori.  To show this, we preform an additional integration with the Toda potential, but now  the initial conditions are taken at different times along the trajectories of the FPU integration, namely, we propagate the initial ensemble with the FPU Hamiltonian for some time $t$ and then switch to the Toda potential and integrate for an additional time of $\Delta t$. The resulting curves are presented in Fig.~\ref{fig:BenettinLike}(b). For these curves, a time-average has been taken as well. Clearly, at different times the FPU dynamics drifts between different Toda tori. This can be thought as an adiabatic theory in standard Hamiltonian dynamics, where the action variables vary slowly in time while the angles rotate rapidly; however, the situation here is not trivial, as we are dealing with an isolated,  many body {\it chaotic} system. One way to interpret this result is along the lines of Ref.~\cite{Lam&Kurchan2014}--- a deterministic quasi-integrable system can be understood as a stochastic one, there, chaos is preliminary restricted on the tori. Indeed, we measured the Lyapunov time in our FPU chain and find it is  $\sim 10^4$, much smaller than the relaxation time.

\section{Weakly-conserved Toda modes in the FPU chain}
\label{sec:TodaJ}

The quasi-static states presented above are characterized by Toda tori, namely, a set of variables which are invariant in the Toda model.  In this section we define the specific choice of such constants of motion, which are used throughout the paper. Then, we discuss their slow motion under the FPU dynamics.

\subsection{Choice of Toda Constants}

The Toda action variables were introduced by Ferguson et. al.~\cite{Ferguson_etal1982}, however, their calculation is intractable. Our choice of invariant quantities is based upon the definition of the exact action variables, capturing their main characteristics. We consider only the case of odd $N$, the case of even $N$ can be readily deduced. Following the analysis of Ref.~\cite{Ferguson_etal1982}, we first give a rigorous procedure for calculating the set of variables, denoted by $\{J_k\}$, and then indicate their essential properties.

\begin{figure}
\centerline{\resizebox{0.45\textwidth}{!}{\includegraphics{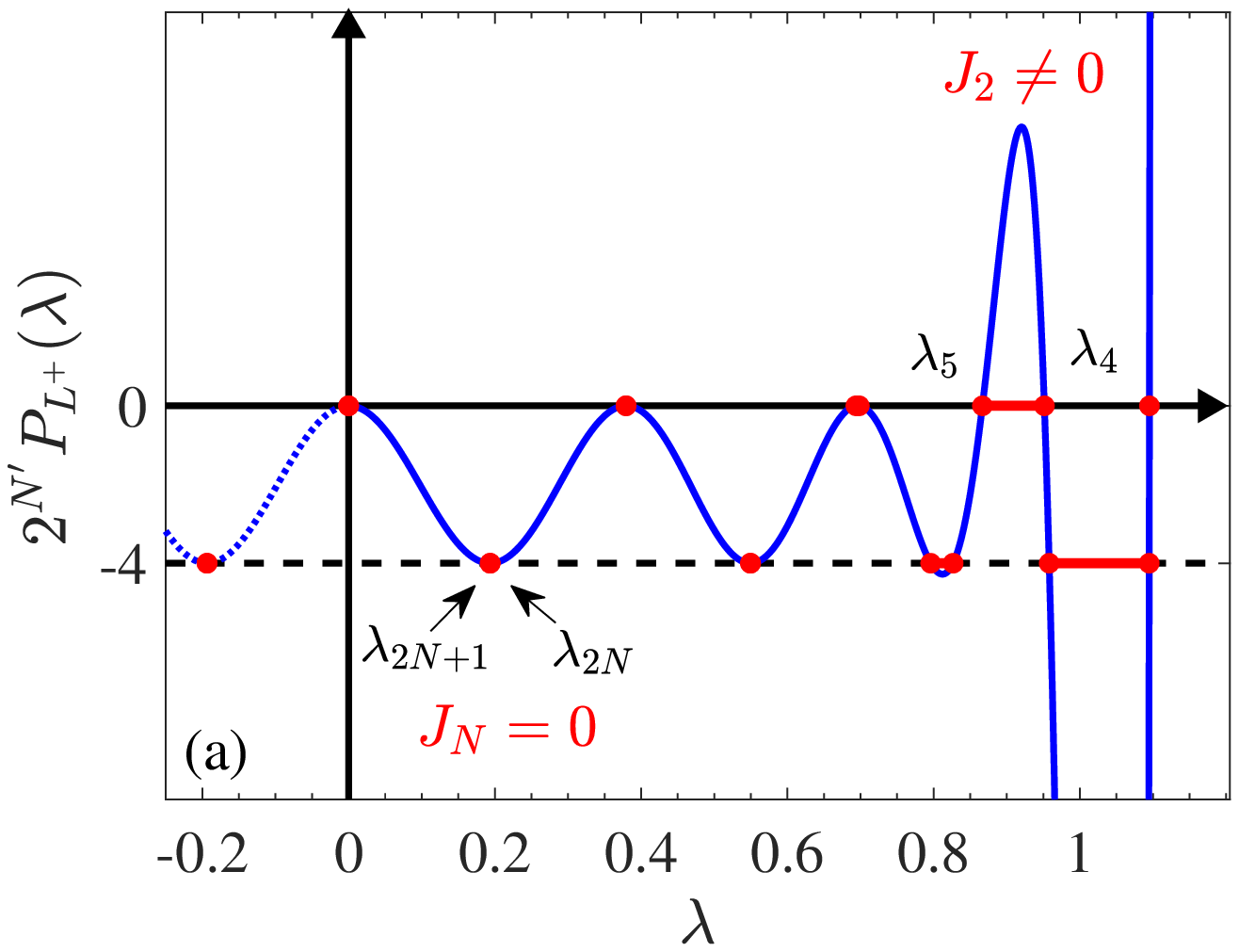}}
\hspace{1cm}
\resizebox{0.45\textwidth}{!}{\includegraphics{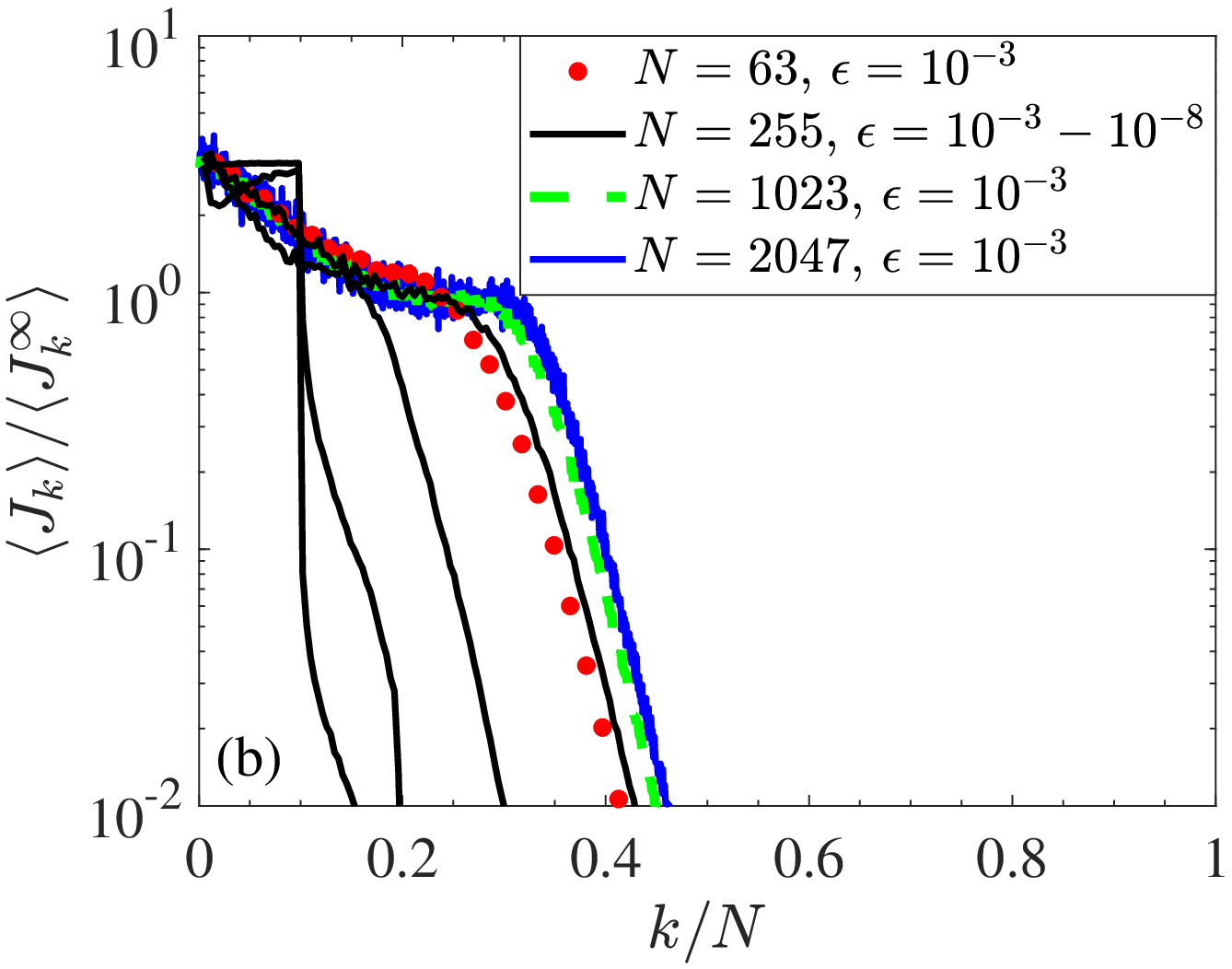}}}
\caption[]{(a) The characteristic polynomial of the matrix $\vec{L}^+$, for a small chain of $N=15$. The eigenvalues of $\vec{L}^{+}$ and $\vec{L}^{-}$ are defined by $P_{L^+}(\lambda)=0$, and $2^{N'}P_L(\lambda)=-4$. The eigenvalues are coming in pairs which are degenerate at zero energy. When the lowest Fourier modes are excited (in the above example only $k=1$ is excited) the degeneracy at the right part of the plot vanishes. For example, the degeneracy of $\lambda_{2N}$ and $\lambda_{2N+1}$ correspond to non-excited Toda mode, whereas the open gap between $\lambda_4$ and $\lambda_5$ correspond to an excited one. (b) The profiles of $\{J_k\}$ for large chains, averaged over  a normal mode ensemble with $K=N/10$. The normalization $\{J^{\infty}_k\}$ corresponds to normal mode ensemble with $K=N$, i.e., a state close to a thermalized chain. The black and colored curves shows the scaling with the energy density $\epsilon$ and system size $N$, respectively.} 
\label{fig:DefJ}
\end{figure}

\begin{itemize}
	\item  \textbf{Defining a periodic chain}--- As a first step we extend the fixed-ends system of $N$ particles into an antisymmetric periodic one with $N'=2(N+1)$ particles:
\begin{equation}
\left\{\begin{array}{l l}
q'_i=q_i,\,\,\,p'_i=p_i, &i=1,\dots,N+1, \\
q'_{N+1+i}=-q_{N+1-i},\,\,\,p'_{N+1+i}=-p_{N+1-i},& i=1,\dots,N+1.	
\end{array}\right.
\label{eq:Per}
\end{equation}
The antisymmetric construction is conserved by the dynamics. As shown below, the extended chain has $N'$ conserved quantities that are degenerated to $N$ quantities for the fixed-ends one.

\item \textbf{Constructing a Lax matrix}--- Next, we define a symmetric matrix $\vec{L}^+$ of size $N'\times N'$,
\begin{equation}
\vec{L}^+=
\begin{pmatrix}
b_1     &a_1     &       &        &        &a_{N'}  \\
a_1     &b_2     &a_2    &        &        &        \\	
        &\ddots  &\ddots &\ddots  &        &        \\	
        &        &\ddots &\ddots  &\ddots  &        \\
        &        &       &a_{N'-2}&b_{N'-1}&a_{N'-1}\\
a_{N'}  &        &       &        &a_{N'-1}&b_{N'}  \\
\end{pmatrix},
\label{eq:Lplus}
\end{equation}
where the unoccupied entries are zero and
\begin{equation}
a_n = \frac{1}{2} e^{\alpha(q'_n-q'_{n-1})},\qquad
b_n = \alpha p'_{n-1}.
\end{equation} 
The eigenvalues of this matrix, $\lambda_n^+$, do not vary in time under the Toda dynamics--- this underlay the integrability of the Toda Hamiltonian, as was proven in earlier works~\cite{Henon1974,Flaschka1974}.

\item \textbf{The characteristic polynomial of $\vec{L}^+$}--- The action variables can be defined through the characteristic polynomial of the Lax matrix $\vec{L}^+$, depicted in Fig.~\ref{fig:DefJ}(a). The polynomial, $P_{L^+}(\lambda)$ , is symmetric about the $y$ axis as a result of the antisymmetric extension to a periodic chain ($\vec{q},\vec{p}\rightarrow\vec{q}',\vec{p}'$). When the chain is at rest, i.e.,  $a_n=1/2$ and $b_n=0$, the extrema of $P_{L^+}(\lambda)$ {\it are degenerate in pairs}, lying on the $x$-axis and the line $y=-4\cdot 2^{-N'}$. The former are the eigenvalues of $\vec{L}^{+}$, and the latter are given by the eigenvalues of another matrix $\vec{L}^{-}$~\cite{Kac&Moerbeke1975}
\begin{equation}
L^-_{ij}=
\left\{\begin{array}{l l}
	L^+_{ij}, & (i,j)\neq (1,N), (N,1)\\ 
	-L^+_{ij}, & (i,j)=(1,N), (N,1)\\
\end{array}\right..
\label{eq:Lminus}
\end{equation}
At finite energy, $\epsilon>0$, the degeneracy is broken (see Fig.~\ref{fig:DefJ}(a)). The action variables are then defined by an integration of a certain complex function
(see \cite{Ferguson_etal1982}) within the corresponding gaps. 
Our strategy here is to consider a much simpler related set of conserved (but not quite `action') variables for the Toda lattice:
they consist of the intervals of these integrals, shown in red in Fig. \ref{fig:DefJ}(a).  These quantities are conserved since the invariance of the eigenvalues of $\vec{L}^+$ impose the invariance of $P_{L^+}(\lambda)$.

\item \textbf{Definition of $\{J_k\}$}--- The aforementioned intervals are given as follows: We calculate the eigenvalues of $\vec{L}^+$ and $\vec{L}^-$ and sort them in an increasing order. The symmetry of the problem yields the following pattern $-\lambda_1, -\lambda_2,\dots -\lambda_{N'-1},0,0,\lambda_{N'-1},\dots,\lambda_2,\lambda_1$, where we recall that $N'=2N+2$.  In the case of odd $N$, the extrama of $P_{L^+}(\lambda)$ lies between the pairs $\{\lambda_{2n},\lambda_{2n+1}\}$. Thus we define define  
\begin{equation}
\{J_k\}^{N}_{k=1}=\{\lambda_{2n}-\lambda_{2n+1}\}^{n=N}_{n=1}.
\label{eq:Jk}
\end{equation}

\item \textbf{Properties}--- While other choices of conserved quantities have simpler definitions, for instance, the trace of $(\vec{L}^+)^p$ with $p=1,\dots,N$~\cite{Henon1974,Flaschka1974}, our choice has several important advantages~\cite{Ferguson_etal1982}: ({\it i}) For small energy, the values of $\{J_k\}$ are correlated with the normal modes of a linear chain--- exciting low/high frequency Fourier modes corresponds to opening gaps (non-degenerated pair of eigenvalues) on the right/left side of the characteristic polynomial; See Fig.~\ref{fig:DefJ}(a). ({\it ii}) gaps which involve $\lambda_n>1$ are related  to excitation of soliton-like waves. ({\it iii}) The action variables of the Toda Hamiltonian are given by integrals of the form $\int_{\lambda_{2n+1}}^{\lambda_{2n}} G(P_{L^+}(\lambda)) d\lambda$, where $G$ is a known function. Thus, the number of degrees of freedom, i.e., non-vanishing action variables, is given by the number of non-vanishing $J_k$.

Since we are interested in the thermodynamics limit, it is crucial to understand how $J_k$ depends on $N$. First, we have the relation 
\begin{equation}
\sum^{N'}_{n=1} (\lambda_n^{+})^{2}=(2\alpha)^2 \mathcal{H}_{\rm Toda}+(N+1),
\label{eq:TodaEnergy1}
\end{equation}
from which we conclude that $\left\{\lambda_n\right\}=O(1)$ in $N$. This implies that $\{J_k\}=O(N^{-1})$. In what follows, we normalized the conserved quantities with $\{J^{\infty}_k\}$, which are the averaged values of $\{J_k\}$ over a normal modes ensemble with $K=N$. The normalization is not necessary, but it provides clearer presentation of the results as $\{J^{\infty}_k\}$ correspond to a situation which resembles equilibration. We  analyzed numerically the behavior of the normal mode ensemble in terms of ${J_k}$. In figure~\ref{fig:DefJ}(b) we show how this ensemble with $K=N/10$ scales with $N$ and $\epsilon$. The scaling $\langle J_k \rangle/\langle J^{\infty}_k \rangle=O(1)$ in $N$ verifies that the chosen quantities are well-behaved thermodynamic intensive variables. The scaling with $\epsilon$ emphasizes the correlation between the Toda and Linear action variables at small energy.  
\end{itemize}

\subsection{Time evolution}
\begin{figure}
\centerline{\resizebox{0.45\textwidth}{!}{\includegraphics{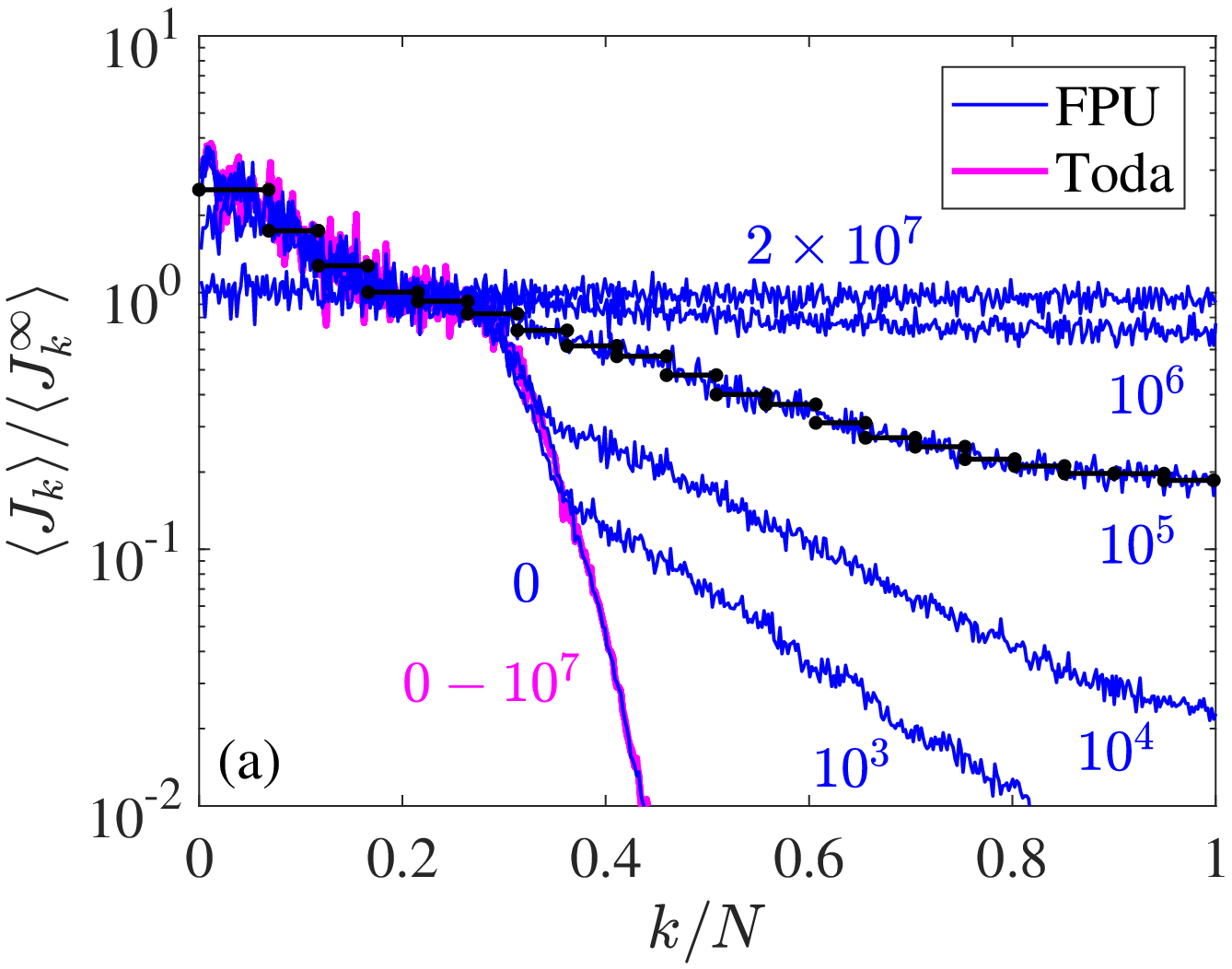}}
\hspace{1cm}
\resizebox{0.45\textwidth}{!}{\includegraphics{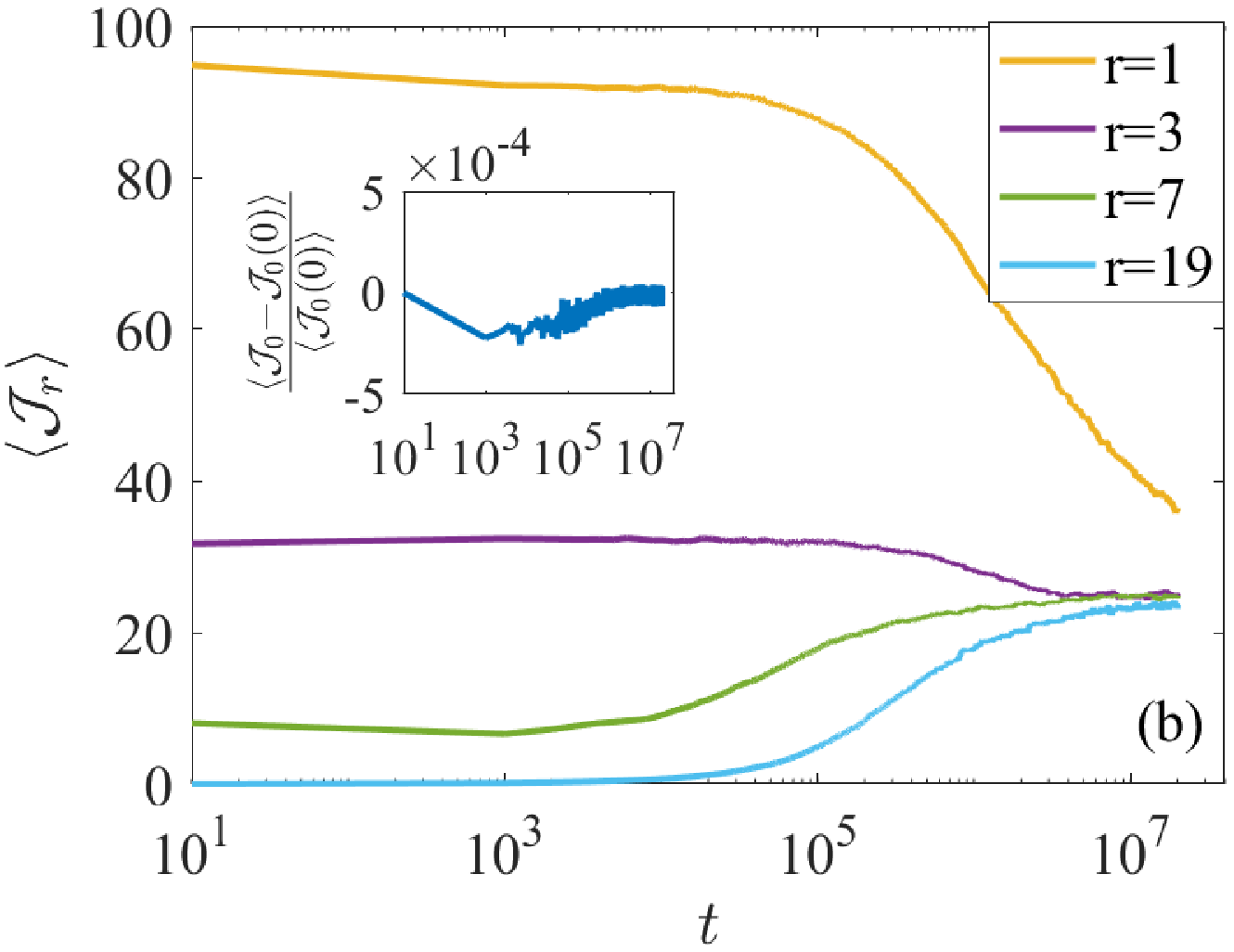}}}
\caption[]{The slow dynamics of the Toda quasi-conserved quantities in the FPU dynamics. (a) The profiles at different times are shown in blue curves. The magenta curves correspond to the Toda dynamics, under which $J_k$ are invariant. The coarse-graining is shown by the binning in black. (b) Examples for the dynamics of the coarse-grained quantities. $\mathcal{J}_1$ correspond to soliton-like waves. The quantities of $\mathcal{J}_r$ with $r>1$ approximately saturate to the same value due to the normalization by $J^{\infty}$ and the equal binning (see Eq.~\eqref{eq:binning}). The inset shows the small variation in time of $\mathcal{J}_0=\mathcal{H}_{\rm Toda}$.} 
\label{fig:JJbar}
\end{figure}

Figure~\ref{fig:JJbar}(a) shows the time-dependent profiles of $\{J_k\}$ along the FPU (blue curves) dynamics. The resemblance of these profiles with the ones of the normal modes' energy in Fig.~\ref{fig:BenettinLike}(a) might be misleading, and a clarification is needed. The two sets of quantities, $\{E_k\}$ and $\{J_k\}$, are essentially different. Along a single trajectory the normal modes fluctuate significantly on short timescales $t<\tau_{\rm Toda}$, where $\tau_{\rm Toda}$ is the typical time to fill the Toda tori. This can be seen by the different profiles of $\langle E_k\rangle$ at $t=0$ and $t=10^3$ in Fig.~\ref{fig:BenettinLike}(a). On the other hand, $\{J_k\}$, being constants of the Toda chain, they vary only on timescales larger than $\tau_{\rm Toda}$. Thus, over time windows of order $\tau_{\rm Toda}$, the average dynamics of the Toda modes underlines the one of the Linear modes and not vice-versa. The resemblance between the profiles originates in the correlation between $\{E_k\}$ and $\{J_k\}$ (at least for large enough $k$), and the averaging over initial conditions.

Finally, let us remark that the magenta curves in Fig.~\ref{fig:JJbar}(a), which correspond to the Toda dynamics, verify that the effects discussed in this paper are not a result of a numerical noise. Such effects are observed for lower variations of $\{J_k\}$, namely, at lower values of the y-axis in Fig.~\ref{fig:JJbar}(a) which are not presented here.


\subsection{Coarse-graining and Generalized Gibbs Ensemble}
\label{ssec:GGE}

The measure of each quasi-static state is defined by the microcanonical distribution $\rho(\vec{x})=\Pi^N_{k=1} \delta(J_k(\vec{x})-J_k)$.  Alternatively, in analogy to Gibbs measure, one may consider the corresponding `canonical' version:
\begin{equation}
\rho_{\rm GGE}(\vec{x})=e^{-\sum \beta_k J_k(\vec{x})}/Z, \qquad Z=\int d\vec{x}e^{-\sum \beta_k J_k(\vec{x})}, 
\label{eq:GGE}
\end{equation}
where $\beta_k$ are Lagrange multipliers; see Ref.~\cite{Yuzbashyan2016} for detailed discussion on such correspondence. This is known as the Generalized Gibbs Ensemble (GGE). Its construction can be understood  as a form  of maximal entropy principle~\cite{Jaynes1957}. Recently, an extensive study addressed GGEs in describing quantum integrable systems~\cite{Vidmar&Rigol2016} (and references therein), for which, e.g., the response and correlation function provide a direct access to $\beta_k$~\cite{Foini_etal2017}. Analogous classical systems has been studied as well~\cite{Cugliandolo2018}. 

In the large $N$ limit, and to the extent that  the distribution of the values of the approximate constant of motion is smooth,
as in Fig. \ref{fig:JJbar},  it is desirable to move to a coarse-grained description: a kind of hydrodynamic limit. 
Such a description guarantees that coarse-grained variables have vanishing fluctuations, and guarantees equivalence of ensembles. The coarse-grained quasi-conserved quantities are defined as 
\begin{equation}
\mathcal{J}_r=\sum^{k_{r+1}}_{k=k_{r}} \left(\frac{J_k}{J^{\infty}_{k}}\right), \quad r=1,2,\dots,R.
\label{eq:binning}
\end{equation} 
By studying different ways for the binning indices, $\{k_{r}\}$, we have found that it is necessary to separate the parts in the profile corresponding to soliton-like modes ($\lambda>1$) from the rest, rather than taking equal spacing in the interval $[1,N]$. In addition, it is crucial to treat separately the set of truly-conserved quantities--- variables which do not vary in the FPU chain, that is, the total energy  ($\mathcal{H}_{\rm FPU}$ in Eq.~\eqref{eq:H_FPU}). (In the case of free ends one should also include the total momentum.) It turns out that in our case one may approximate energy by  its Toda value, as this quantity changes little over the whole range of times; see inset in Fig.~\ref{fig:BenettinLike}(b). Thus, using Eq.~\eqref{eq:TodaEnergy1} we define
\begin{equation}
\mathcal{J}_0\equiv \mathcal{H}_{\rm Toda}=\frac{2\sum^{N+1}_{n=1} (\lambda_n^{+})^{2}-(N+1)}{(2\alpha)^2}.
\label{eq:Jbar0}
\end{equation}
Fig.~\ref{fig:JJbar}(b) shows the dynamics of different quasi-conserved, coarse-grained quantities.

To demonstrate that the coarse-grained set $\{\mathcal{J}_r\}^R_{r=0}$ captures the statistics of the quasi-static states at some time $t$ we preform a micro-canonical sampling of $\rho(\vec{x})=\Pi^R_{r=0} \delta\left(\mathcal{J}_r(x)-\langle\mathcal{J}_r\rangle(t)\right)$ as follows: We start with a random initial condition $\vec{x}_0$, then we propose a random change $\Delta\vec{x}$ and accept it only if the function $\sum_r\left(\mathcal{J}_r(x)-\langle\mathcal{J}_r\rangle(t)\right)^2$ decreases. Between the random changes in $\vec{x}$ we run the Toda dynamics for some random interval of time. This allows large changes in the chain without changing $\mathcal{J}_r$, which leads to a better convergence of the minimization procedure. In figure~\ref{fig:BenettinLike}(b) we illustrate the micro-canonical sampling for $t=10^5$ and $10^6$, by plotting the expectation values of $E_k$ over the resulting ensemble. Clearly, the set $\{\mathcal{J}_r\}$ represents the statistics well.

Below we discuss two aspects of the GGE description--- how it underlies a fast numerical integration and a fluctuation theorem.

\section{Fast Numerical Scheme}
\label{sec:numerical}

\begin{figure}
\centerline{\resizebox{1\textwidth}{!}{\includegraphics{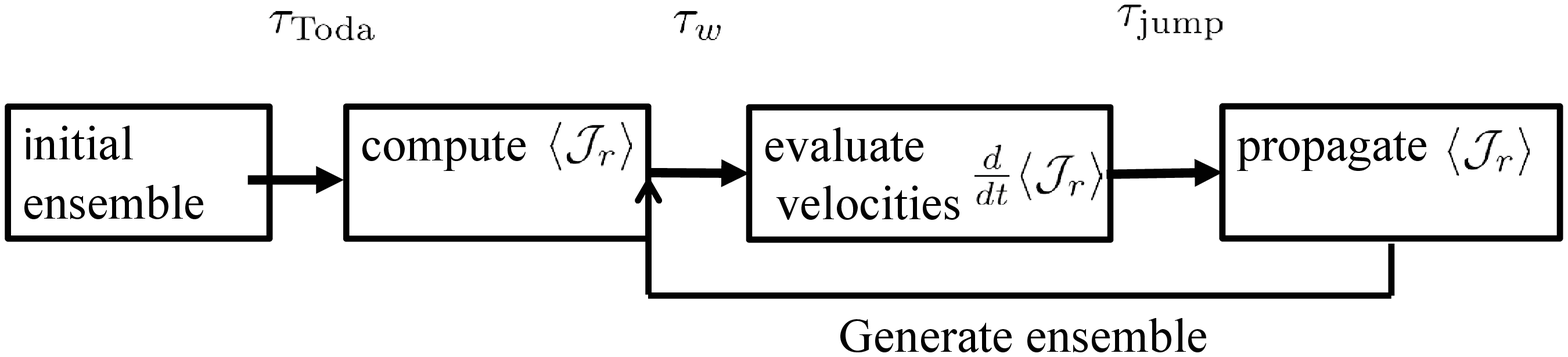}}
}
\caption[]{The indirect numerical protocol.} 
\label{fig:diagram}
\end{figure}

The self-averaging, slow dynamics evident in figure~\ref{fig:JJbar}(b) suggests that it is sufficient to follow the dynamics of $\langle \mathcal{J}_r\rangle$ in order to integrate the complete ensemble. Unfortunately, the dynamical equations for the weakly-conserved quantities cannot be written in a form which is analytically manageable. However, we can devise a fast numerical scheme as an indirect integration of their trajectories (see also diagram in Fig.~\ref{fig:diagram}):
\begin{enumerate}
	\item Start at $t=0$ with a normal mode ensemble.
	\item Perform a direct integration of the system for some transient time, $\tau_{\rm Toda}$, after which the initial tori are completely explored (i.e., $\tau_{\rm Toda}=10^3$ in the example in Fig.~\ref{fig:BenettinLike}(a)). 
	\item Estimate the velocities $\frac{d\langle \mathcal{J}_r\rangle}{dt}$. This can be done by an additional direct integration for some time window $\tau_{w}>\tau_{\rm Toda}$, from which we have 
$$
\frac{d\langle \mathcal{J}_r\rangle}{dt}\approx \frac{\langle \mathcal{J}_r\rangle(t+\tau_w)-\langle \mathcal{J}_r\rangle(t)}{\tau_w}.
$$
\item Extrapolate linearly the values $\langle \mathcal{J}_r\rangle$ to a much larger time $\tau_{\rm jump}>\tau_{w}$
$$
\langle \mathcal{J}_r\rangle (t+\tau_{\rm jump})\approx \langle \mathcal{J}_r\rangle(t)+\tau_{\rm jump}\frac{d\langle \mathcal{J}_r\rangle}{dt}.
$$
This is justified by the slow dynamics of the quasi-conserved quantities. 
\item Generate a new ensemble with $\langle \mathcal{J}_r\rangle (t+\tau_{\rm jump})$; see details below. 
\item Iterate again, going back to step 3.
\end{enumerate}
This numerical method falls into the class of Heterogeneous Multiscale Methods~\cite{E2007}. Note that {\em the coarse-grained variables do not fluctuate: the situation is just like integrating hydrodynamic equations}. The implementation of the scheme relies on the time-scale separation $\tau_{\rm Toda}\ll \tau_{w}\ll \tau_{\rm jump}$. (We may  relax the first inequality $\tau_{\rm Toda}\ll \tau_{w}$ by running in parallel an ensemble of dynamics and averaging; however, we have found the fluctuations of averages are too large, for a reasonable amount of copies.)

 The key point of the scheme is to efficiently generate sets of initial conditions within  coarse-grained tori, given the values of  $\langle \mathcal{J}_r\rangle$. This can be done by the zero-temperature minimization described at the end of Sec.~\ref{sec:TodaJ}. Alternatively, we can use a Conjugate-Gradient minimization scheme~\cite{NumericalRecipes} starting with an ensemble of random initial points. During the down-hill steps we run a direct integration of the Toda dynamics, for time intervals $\gamma\cdot\tau_{\rm Toda}$, with some random number $\gamma$, allowing a better sampling of the tori and better mixing. The function we minimize is 
\begin{equation}
f_{\min}(\vec{x})=\frac{\left(\mathcal{J}_0(\vec{x})-\langle\mathcal{J}_0\rangle \right)^2}{C \langle\mathcal{J}_0\rangle}+ \sum^R_{r=1} \frac{\left(\mathcal{J}_r(\vec{x})-\langle\mathcal{J}_r\rangle \right)^2}{\langle\mathcal{J}_r\rangle}, 
\end{equation}
where the parameter $C$, chosen to be $0.01$, gives a more strict restriction on the truly-conserved quantity. Clearly, the slower the dynamics (smaller $\epsilon$), and the smaller the fluctuations (larger $N$), the more beneficial is the protocol.  

\begin{figure}
\centerline{\resizebox{0.5\textwidth}{!}{\includegraphics{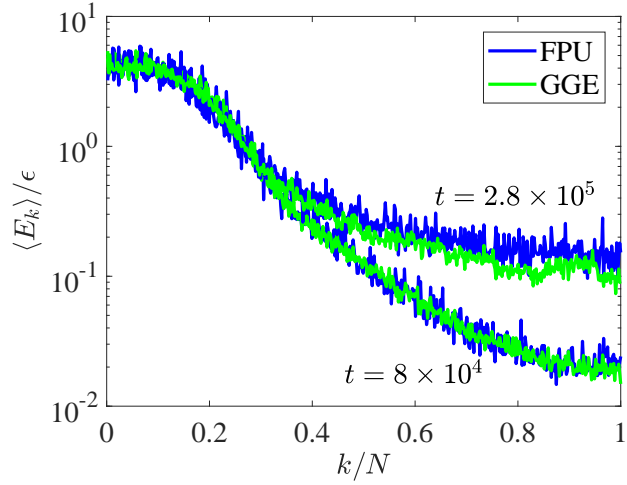}}
}
\caption[]{A demonstration of the numerical scheme presented in Sec.~\ref{sec:numerical}. We employ three jump iterations to get from $t=8\times 10^4$ to $t=2.8\times 10^5$.} 
\label{fig:numerical}
\end{figure}

In figure~\ref{fig:numerical} we show a proof of concept of this scheme, implementing 3 iterations to reach $t=2.8\times 10^5$, using an ensemble of 100 samples. The parameters we use, $N=511$ and $\epsilon=10^{-3}$, give a marginal time-scales separation, and thus, do not give an explicit proof that the indirect scheme is faster than a simple integration. Nevertheless, this example provides us estimates of computational times, from which we can quantify the feasibility of the suggested protocol.

Let us define $T_{\rm direct}$ as the computation time of one FPU integration with time-step $\Delta t$, $T_{\min}$ as the computation time for minimization, and $N_{\rm samp}$ is the number of samples {\it per CPUs} (as the indirect scheme can be simply parallelized). To evaluate the efficiency of the protocol we compare the computation time of indirect iteration, from time $t$ to $t+\tau_{\rm jump}$:
$$
\left[T_{\rm direct}(\tau_{w}/\Delta t)+T_{\min}\right]\times N_{\rm samp}.
$$
with the computation time of a direct integration:
$$
T_{\rm direct}(\tau_{\rm jump}/\Delta t).
$$
Note that we are comparing a direct integration of a single initial condition with an indirect one of an ensemble. The indirect scheme supersedes the direct one for jump-steps satisfying 
\begin{equation}
\tau_{\rm jump}>\left[\tau_{w}+\frac{T_{\min}}{T_{\rm direct}}\Delta t\right]\times N_{\rm samp}.
\end{equation}
For the specific example in figure~\ref{fig:numerical} we use a time-step of $\Delta t=0.1$ and find that $T_{\rm direct}\approx 0.002$ seconds, whereas on average $T_{\min}=100$ seconds. Therefore, the indirect scheme has an advantage for jump-steps $\tau_{\rm jump}>10^4N_{\rm samp}$. Note that $\tau_{\rm jump}$ is bounded by how slow the quasi-conserved quantities are, namely, on what time-scales a linear extrapolation with $\frac{d\langle \mathcal{J}_r\rangle}{dt}$ is valid. Technical details related to the numerical procedures are given in the Appendix.

\section{Generalized Exchange Fluctuation Theorem}
\label{sec:GXFT}

Let us now return to the GGE description of a quasi-static state. 
\begin{equation}
\rho_{\rm GGE}(\vec{x})=e^{-\sum^R_{r=0} \tilde{\beta}_r\mathcal{J}_r(\vec{x})}/Z,
\label{eq:GGEbar}
\end{equation}
where $\tilde{\beta}_r$ correspond to the coarse-grained variables. The directed flow of $\{\mathcal{J}_r\}$ toward equilibration, as induced by the breaking of integrability, involves fluctuations  which are expected to vanish in the limit $R \to \infty$ (cfr Eq (\ref{eq:binning})) after $N\to\infty$ (see Fig.~\ref{fig:JJbar}(b)). 

For a finite system and finite times, there is a Fluctuation Theorem (FT) which describes the probability of inverse flows.  In Ref.~\cite{Goldfriend&Kurchan2018} we derived the FT and discussed its consequences, which we now summarize.
We consider the change $\Delta \mathcal{J}_r$ after time $t$ by the quantity $u\equiv \sum^R_{r=0} \tilde{\beta}_r\Delta \mathcal{J}_r $, where an initial GGE state with corresponding $\tilde{\beta}_r$ is assumed.  The FT states that the probability of the system to assume a value $u$, designated by $P(u)$, satisfies 
\begin{equation}
\frac{P(u)}{P(-u)}=e^u.
\label{eq:GXFT}
\end{equation}
The large-deviations described by Eq.~\eqref{eq:GXFT} depend on how the system is close to integrability, that is, for an integrable system, $u=0$ identically. Therefore, unlike usual FTs neither the system size, nor the time, need to be small for the relation to be relevant, provided that the departure from integrability is small.

\begin{figure}
\centerline{\resizebox{0.5\textwidth}{!}{\includegraphics{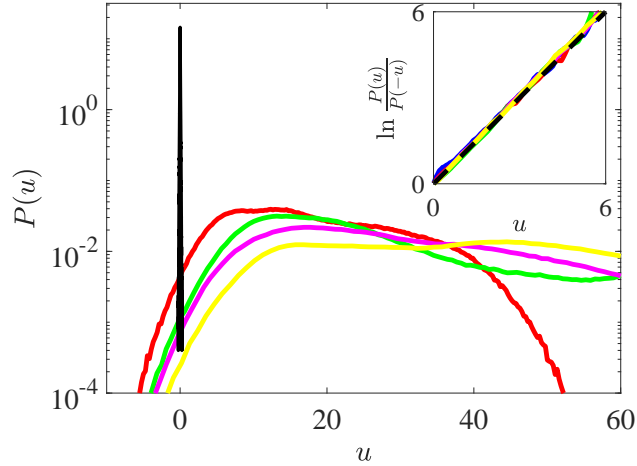}}
}
\caption[]{Illustration of the generalized exchange fluctuation theorem for a FPU chain, with $N=15$. We generate an initial GGE, $\rho(\vec{x})\propto \exp\left[-\sum\beta_k J_k(\vec{x})\right]$, for some chosen set of $\{\beta_k\}$. Then, we propagate the system from $\vec{x}$ to $\vec{x}^t$ for some time $t$. Main: colored--- probability distribution of $u=\sum_k \beta_k(J_k(\vec{x}^t)-J_k(\vec{x}))$ at different times when the system evolves with the FPU Hamiltonian. black--- the distribution for the Toda dynamics, corresponding solely to random numerical error. Inset: verification of the fluctuation theorem (Eq.~\eqref{eq:GXFT}).} 
\label{fig:GXFT}
\end{figure}

For large systems, the assumption of initial GGE state can be relaxed. If the hydrodynamic limit holds, e.g., in the initial normal modes ensemble studied in this paper, equivalence of ensembles guarantees that Eq.~\eqref{eq:GXFT} remains intact up to small corrections; see Ref.~\cite{Goldfriend&Kurchan2018}. The slow drift of the coarse-grained quantities $\{\mathcal{J}_r\}$ toward equipartition induces a slow drift of the GGE and the Lagrange multipliers $\tilde{\beta}_r$ accordingly. As $t \rightarrow \infty$ the quasi-conserved quantities, $\mathcal{J}_r$ with $r=1,\dots,R$, reach a constant value an no longer impose any constraint, i.e., $\tilde{\beta}_r \rightarrow 0$ at long times. On the other hand, the truly-conserved quantity $\mathcal{J}_0$ and the corresponding Lagrange multiplier $\tilde{\beta}_0$ reaches a constant value as $t\rightarrow\infty$ (for this one should take the exact $\mathcal{J}_{0}=\mathcal{H}_{\rm FPU}$ and not its approximated value $\mathcal{H}_{\rm Toda}$ used in the simulations).

From Fig.~\ref{fig:JJbar}(b) we can see that at late times, $t\sim 10^7$, all the Toda modes starts to saturate, except the one which is related to solitons. Thus, at long times, the FT concerns the ratio of probability of death to resurrection of solitons. 

A verification of the FT is presented in Fig.~\ref{fig:GXFT} for a small system of $N=15$. For this example we do not use the normalized coarse-gained quantities,  but rather take the original set $\{J_k\}$ defined in Sec.~\ref{sec:TodaJ}.


\section{Conclusions}
\label{sec:Con}

We have studied the slow evolution of the FPU chain, and established the quasi-static motion in the space of weakly-conserved quantities, the latter corresponding to the integrable Toda chain. We have shown how this picture opens a window to explore the system more efficiently. First, we have offered a fast numerical scheme for integrating the quasi-integrable chain. Secondly, based on the slowly evolving GGE, the motion within the entropy-landscape has been characterized by a Fluctuation Theorem.  

In this work we have focused on the specific nonlinear FPU chain; nevertheless, the picture presented here should be  useful for quasi-integrable systems in general. Going back to the definition given in the Introduction, quasi-integrability refers to systems with time-scale separation resulting in slow dynamics of weakly-conserved quantities. In particular, the numerical scheme introduced in Sec.~\ref{sec:numerical} can be generalized to quasi-integrable systems of few degrees of freedom, e.g., the Solar System. In such cases,  fluctuations along the drift of the weakly-conserved quantities are expected to be significant. Therefore, the modification of the scheme includes evaluation of a diffusive motion, that is, the extrapolation step is now a Langevin dynamics, $\frac{d}{dt}\langle \mathcal{J}_r\rangle \approx \mathcal{V}_r+\eta_r(t)$, where $\mathcal{V}_r$ is the evaluated constant drift velocity and $\eta_r$ is uncorrelated noise of variance $D_{rs}=\langle \mathcal{J}_r\mathcal{J}_s\rangle$. 

As an example, let us consider the analysis of the Solar System dynamics in the spirit of our study on the FPU chain. A relevant numerical experiment is the one which was preformed by Laskar~\cite{Laskar2008}: we integrate the secular equations for the solar system with an initial ensemble concentrated around the present state of the 8 planets. Then, along the dynamics we can measure observables such as the probability distribution function (PDF) for the eccentricities or inclination angles of a planet. In analogy to Fig.~\ref{fig:BenettinLike}, we can ask whether there are weakly-conserved quantities that determine these PDFs, and what is their dynamics. A good candidate for the underlying integrable model is the Laplace-Lagrange Hamiltonian, which corresponds to 16 action-angle variables $\{I_r,\theta_r \}^{16}_{r=1}$ (in the secular equations one averages over rotations around the sun, eliminating one degree of freedom per planet~\cite{Morbidelli}). Recently, it  was shown that the PDFs for the outer planets, which are stationary for 5 Gyrs, are determined by the microcanonical ensemble $\Pi_r\delta(I_r-I^0_r)$, where $I_r^0$ are the initial values~\cite{Mogavero2017}. On the other hand, the PDFs of the inner planets are spreading with time. This time-evolving effect should result not only from a  deterministic drift of $I_r$, but rather from the diffusion of the initial ensemble--- that is, any element of the initial ensemble follows a slow, but different, trajectory, which at short time windows can be determined by some Laplace-Lagrange torus.

In the realm of quantum mechanics, the study of Generalized-Gibbs-Ensembles has recently attracted a lot of attention, where the main program is the understanding of the  relaxation of systems after quenches. Experiments on cold atoms allow to explore the properties of isolated systems. When the system is integrable, it cannot relax to the usual Gibbs measure but expectation values of observables can be computed with a GGE; see original experiments and theory in Refs.~\cite{Kinoshita_etal2004} and~\cite{Rigol_etal2007} respectively, or Ref.~\cite{Calabrese2016} for recent comprehensive collection of reviews. In the case of quasi-integrable systems, there is a long-lived transient state, referred to as  {\it prethermalized phase}, in which the system can be described by a GGE before it eventually relaxes to the Gibbs measure. In analogy to the classical systems discussed here, one may think of prethermalization as the motion within Hilbert subspaces with similar quantum numbers for the quasi-conserved quantities, the quantum analogue of our  Toda tori. The ideas we have presented in Secs.~\ref{sec:numerical} and~\ref{sec:GXFT} should hence be applicable to quasi-integrable quantum systems. Indeed, a similar Fluctuation Theorem can be derived~\cite{Hickey&Genway2014,MurPetit_etal2018,Goldfriend&Kurchan2018}. Although usually numerical simulations of quantum models cannot address large systems, the numerical scheme we have devised in Sec.~\ref{sec:numerical}, or its extension presented in the previous paragraph, should be found useful to explore the slow relaxation of quantum isolated systems. We note that for some quantum chains, a kinetic theory describing time-evolving GGE can be derived and solved for the slow dynamics toward equilibration~\cite{Stark&Koller2013,Bertini_etal2015}.

Going back to the relation between the Toda and FPU chains, our choice of conserved quantities allows us to explore the whole spectrum of Toda modes and provides a well-defined macroscopic variables. This is in contrast with previous studies e.g., Refs.~\cite{Ponno_etal2011,Benettin_etal2013}, where a different choice of Toda constants was considered, $I_p={\rm Tr} ((\vec{L^+})^p)$, with $p=2,4\dots,2N+2$. These are the original integrals of motion that  proved the integrability of the Toda chain~\cite{Flaschka1974,Henon1974}. Clearly, higher moments with $p\sim N$ are not well-defined thermodynamic variables as they scale as $O(N^N)$. Indeed, we have found out that an implementation of our numerical scheme with a coarse-grained version for the set $\{I_p\}$, or $\{\lambda_k^+\}$ (the eigenvalues of $\vec{L^+}$) performs very badly. One explanation of this failure is that these sets are biased by the larger eigenvalues of $\vec{L^+}$, with $\lambda_k^+>1$, which are related to soliton-like waves. We note that the choice of conserved quantities for the GGE depends on the observables one wishes to measure: here we have shown that the  set $\{\mathcal{J}_r\}$ is appropriate for the  study of the slow equilibration dynamics.  

As we have mentioned in the Introduction, our approach to deterministic quasi-integrable systems is inspired by the behavior of integrable systems with stochastic perturbation, as analyzed in Ref.~\cite{Lam&Kurchan2014}. We would wish  to rationalize the FPU dynamics as a Toda chain perturbed by (an internally-generated) noise. In particular, we have found that  numerical  noise   in the integration of the Toda chain brings about effects that are similar to those of the integrability-breaking  FPU ones: we observe this in the hidden lower parts of Fig.~\ref{fig:JJbar}(a), where the  Toda (magenta) curves do change in time as a result of  the numerical noise, an effect that becomes  weaker when the integration step of the simulation is reduced.

\section{ACKNOWLEDGMENTS}
We thank Freddy Bouchet, Federico Mogavero, and Jacopo de Nardis for fruitful discussions. TG and JK are supported by the Simons Foundation Grant No. 454943.

\appendix

\section{Numerical procedures}

The numerical computations in the current paper were done with Matlab version R2017b. For integrating the Hamilton equations for the non-linear chains, we use a 4-th order St{\"o}rmer-Verlet method according to Yoshida scheme~\cite{Yoshida1990}, with a time-step of 0.1. As mentioned in the Sec.~\ref{sec:Con}, we have also checked finer time-steps. These result in increasing accuracy of the Toda constants over time (when integrating the Toda equations), but do not affect the FPU behavior studied here.

The most complex part in our numerical computations is the analysis of the Toda constants of motion. In this Appendix we provide more details on the construction of  these quantities, as well as of their derivatives that are used for the Conjugate-Gradient in Sec.~\ref{sec:numerical}. 

\subsection{Constructing $\{J_k\}^N_{k=1}$}

The matrices $\vec{L}^+$ and $\vec{L}^-$ are sparse, however, calculating their eigenvalues is numerically expensive. One way to decrease the computation time is by transforming these almost tridiagonal matrices to an exact pentadiagonal ones, following the procedure in Ref.~\cite{Ferguson1980}. Given the matrix $\vec{L}^+$ in Eq.~\eqref{eq:Lplus}, we define the symmetric pentadiagonal matrix
\begin{eqnarray} 
\vec{d}_{0}&=&\{b_1,b_{N'},b_2,b_{N'-1},\dots,b_{N'-1},b_{N'/2+2},b_{N'/2},b_{N'/2+1}\}, \\
\vec{d}_{\pm 1}&=&\{a_{N'},0,\dots,0,a_{N'/2}\},\\
\vec{d}_{\pm 2}&=&\{a_1,a_{N'-1},a_2,a_{N'-2},\dots,a_{N'/2-2},a_{N'/2+2},a_{N'/2-1},a_{N'/2+1}\}, 
\end{eqnarray}
where $\vec{d}_{0}$ is the main diagonal and $d_{\pm m}$ is the $m$ diagonal above or below the main one. The matrix that correspond to $\vec{L}^{-}$ is defined in a similar way. The pentadiagonal matrices, designated by $\vec{M}^{\pm}$, are given by a permutation transformation of $\vec{L}^{\pm}$ respectively, and thus have the same eigenvalues.

Once the eigenvalues of $\vec{L}^+$ and $\vec{L}^-$ are calculated, they are sorted and the construction of $\{J_k\}^{N}_{k=1}$ is given by Eq.~\ref{eq:Jk}.

\subsection{Derivatives of $\{J_k\}^N_{k=1}$}

The core of this calculation is the evaluation of the gradient of the eigenvalues of $\vec{M}^{\pm}$.  Consider a matrix $\vec{A}(\vec{x})$ with eigenvectors $\vec{v}_n(\vec{x})$ and eigenvalues $\nu_n(\vec{x})$. First, we can write
$$
A(\vec{x}+h \hat{x}_k)\approx A(\vec{x})+h\frac{\partial \vec{A}(\vec{x})}{\partial x_k},
$$
then, from first order perturbation theory we know that the eigenvalues of $\vec{A}(\vec{x}+h\hat{x}_k)$ reads
$$
\nu_n(\vec{x}+h\hat{x}_k)=\nu_n(\vec{x})+h\cdot \vec{v}^T_n(\vec{x})\cdot\frac{\partial \vec{A}(\vec{x})}{\partial x_k} \cdot \vec{v}_n(\vec{x}).
$$
This implies that 
\begin{equation}
\frac{\partial \nu_n(\vec{x})}{\partial x_k}= \vec{v}_n^T(\vec{x})\cdot \frac{\partial \vec{A}(\vec{x})}{\partial x_k}\cdot\vec{v}_n(\vec{x})
\label{eq:gradnu}
\end{equation}

From Eq.~\eqref{eq:gradnu} we learn that in order to evaluate $\nabla \lambda_n^{\pm}$ one must compute the eigenvectors of $\vec{L}^{\pm}$ (or $\vec{M}^{\pm})$, as well as the derivatives $\frac{\partial \vec{L}^{\pm}(\vec{x})}{\partial x_k}$. The latter can be calculated explicitly. The former procedure takes much more cpu-time than just computing the eigenvalues, however, we note that the sparseness of the matrices facilitates the tensor products in Eq.~\eqref{eq:gradnu}. 


\bibliography{QuasiIntegrableFPU}

\end{document}